\documentclass[twocolumn,showpacs,preprintnumbers,amsmath,amssymb]{revtex4}
\begin{document}
\title{The quantum bit commitment: a finite open system approach for 
a complete classification of protocols}
\author{Giacomo Mauro D'Ariano}
\email{dariano@unipv.it}
\affiliation{{\em Quantum Optics and Information Group} of the INFM,
unit\`a di Pavia}
\affiliation{Dipartimento di Fisica ``A. Volta'', via Bassi 6, I-27100 Pavia, Italy}
\homepage{http://www.qubit.it}
\affiliation{Department of Electrical and Computer Engineering,
Northwestern University, Evanston, IL  60208}
\date{September 26, 2002}
\begin{abstract}
Mayers \cite{MayersQBC97}, Lo and 
Chau\cite{LoChauQBC97,LoQBC97} argued that all 
quantum bit commitment protocols are insecure, because there 
is no way to prevent an Einstein-Podolsky-Rosen (EPR) cheating 
attack. However, Yuen \cite{yuen12capri,yuenqbc3,yuenqbc4} 
presented some protocols which challenged the previous 
impossibility argument. Up to now, it is still debated whether 
there exist or not unconditionally secure protocols\cite{yuenqbclast}. 
In this paper the above controversy is addressed. For such purpose, a 
complete classification of all possible bit
commitment protocols is given, including all possible cheating
attacks. Focusing on the simplest class of protocols 
(non-aborting and with complete and perfect verification), it is 
shown how naturally a game-theoretical situation arises.
For these protocols, bounds for the cheating probabilities 
are derived, involving the two quantum operations encoding the 
bit values and their respective alternate Kraus decompositions. 
Such bounds are different from those given in the impossibility 
proof\cite{MayersQBC97,LoChauQBC97,LoQBC97}. The whole classification
and analysis has been carried out using a {\em finite open system}
approach. The discrepancy with the impossibility proof is 
explained on the basis of the implicit adoption of a {\em closed
system approach}---equivalent to modeling the commitment as performed
by two fixed machines interacting unitarily in a overall {\em closed
system}---according to which it is possible to assume as {\em openly
known} both the initial state and the probability distributions for all secret
parameters, which can be then {\em purified}. This approach 
is also motivated by existence of unitary extensions for any open
system. However, it is shown that the closed system approach for the
classification of commitment protocols unavoidably leads to 
infinite dimensions, which then invalidate the continuity argument at
the basis of the impossibility proof.
\end{abstract}
\maketitle
\section{Introduction}
Among all kinds of quantum cryptography protocols, the
quantum bit commitment is a crucial element to build up more 
sophisticated protocols, such as remote quantum gambling
\cite{gambl}, coin tossing  \cite{BCJL}, and unconditionally 
secure two-party quantum computation \cite{LoPop}.
Therefore, it is of practical relevance to establish if there exist
secure quantum bit commitment protocols. \par 
In the bit commitment Alice provides Bob with a piece of evidence that
she has chosen a bit $b=0,1$ which she commits to him. Later, Alice
will open the commitment, revealing the bit $b$ to Bob, and proving
that it is indeed the committed bit with the evidence in Bob's possession.  
Therefore, Alice and Bob should agree on a protocol which satisfies
simultaneously the three requirements: (1) it must be {\em
concealing}, namely Bob should not be able to retrieve $b$ before the
opening;  (2) it  must be {\em binding}, namely Alice should not be
able to change $b$ after the commitment; (3) it must be
{\em verifiable}, namely Bob must be able to check $b$ against the
evidence in his possession, according to the rules of the 
protocol. In a in-principle proof of security of the commitment it is
supposed that both parties possess unlimited technology, e. g. 
computational power, space, time, etc., and the protocol is said   
{\em unconditionally secure} if neither Alice nor Bob can cheat with
significant probability of success as a consequence of physical laws.

In 1993, a quantum mechanical protocol was proposed\cite{BCJL}, and
the unconditional security of this protocol has been generally
accepted for long time. The insecurity of this protocol was shown by 
Mayers, Lo and Chau\cite{MayersQBC97,LoChauQBC97,LoQBC97} in 1997,
where it was recognized the possibility for Alice to cheat by entangling the
committed evidence with a quantum system in her possession, and it was argued
that no unconditionally secure protocol is possible. Finally after
2000 Yuen\cite{yuen12capri,yuenqbc3,yuenqbc4} presented some protocols
which challenged the previous impossibility proof, mostly on the basis
of the possibility of encoding the bit on an {\em anonymous state}
given to Alice by Bob and known only to him, and suggesting
the use of {\em decoy systems} that make the protocol concealing in
the limit of infinitely many systems, with the possibility for Bob of
performing his quantum measurement before Alice opening, whence
disputing the general availability of EPR cheating for Alice. 
Besides the above schemes, protocols have also been suggested
based on the theory of special relativity \cite{kent}
(for historical reviews on the quantum bit commitment see 
Refs. \cite{LoQBC97,yuenqbc4}). Here, however,  we will consider 
only non relativistic protocols. 

In this paper, in order to provide clarifications on the issue of 
existence of unconditionally secure protocols, we will give a complete 
classification of all possible bit commitment protocols based on a 
single commitment step, and show how a multi-step commitment 
can be reduced to a single one. We will analyze all possibilities of 
cheating for both parties. Then, we will focus on the simplest 
class of protocols, namely the non-aborting protocols with 
complete and perfect verification, showing that naturally a
game-theoretical situation arises.  As we will see, even though
perfectly concealing protocols are certainly not binding (i. e.  Alice
has a unit cheating probability),  the protocol could still be binding
if it is $\varepsilon$-concealing. 
Bounds for the cheating probabilities of these protocols are
derived, involving the two quantum operations encoding the 
bit values and their respective alternate Kraus decompositions. 
Such bounds turn out to be different from those given in the 
impossibility proof  \cite{MayersQBC97,LoChauQBC97,LoQBC97}.
In the final discussion we will see how 
the discrepancy between the two opposite analysis arises, as a result
of the restrictive assumption---which lies beneath the impossibility 
proof---of openly known, whence purifiable, probability 
distributions for all secret parameters. Such an assumption
is equivalent to {\em modeling} the commitment as a {\em closed}
system made of two fixed machines interacting unitarily. It is shown
that such modeling, along with the requirement of unlimited
technology, necessarily lead to infinite dimensions,  which invalidate the
continuity argument at the basis of the impossibility proof. 
Instead, one either needs to prove the continuity argument for infinite
dimensions, or else must adopt a {\em finite open system} approach.
\section{The classification of protocols}
The most general bit commitment scheme with a single step is of the 
form: (1) Bob prepares the Hilbert space ${\sf H}$ with the {\em anonymous
state} $|\varphi{\rangle}\in{\sf H}$, and sends ${\sf H}$ to Alice; (2) Alice {\em
modulates} the value $b$ of the committed bit on the anonymous state 
$|\varphi{\rangle}$ and sends the output back to Bob. The {\em bit
modulation} is a quantum operation (QO) parametrized by $b=0,1$. 
It is clear that this general scheme contains all possibilities,
including the anonymous-state based protocols of Yuen
\cite{yuen12capri,yuenqbc3,yuenqbc4,yuenqbclast}, and, as a special
case, the original protocols by Mayers \cite{MayersQBC97}, and Lo and  
Chau\cite{LoChauQBC97,LoQBC97}, which correspond to 
{\em openly known} state $|\varphi{\rangle}$. 

\subsection{Bit modulation} 
To make the protocol {\em concealing} and at the
same time {\em verifiable}, the modulation must be a choice between
two ensembles of QO's $\{{\mathrm{M}}_j^{(b)}\}$ for
$b=0,1$, from ${\sf T}({\sf H})$ to ${\sf T}({\sf K})$, where ${\sf T}({\sf H})$ 
denotes the set of
traceclass operators on ${\sf H}$, and generally the two Hilbert
spaces ${\sf K}$ and ${\sf H}$ are not isomorphic. We will name the cases 
${\sf K}\supseteq{\sf H}$ and ${\sf K}\subseteq{\sf H}$ {\em extending modulation} 
and {\em restricting modulation}, respectively, the extending case
including the possibility of using {\em decoy systems}. The variable
$j$ is a {\em secret parameter} known only to Alice, parametrizing the
choice of different forms for the modulation, and which will be declared
to Bob at the opening. 

\subsection{The secret-parameter space} 
Alice has always the option of
choosing $j$ by preparing a {\em secret-parameter space ${\sf P}$} in a
state chosen from an orthonormal set $\{ |j{\rangle}\}$, and performing the
QO on ${\sf H}\otimes{\sf P}$
\begin{equation}
\mathbb{M}^{(b)}\doteq
\sum_j{\mathrm{M}}_j^{(b)}\otimes\mathrm{P}_j,
\end{equation}
with $\mathrm{P}_j$ representing the orthonormal projection map
$\mathrm{P}_j(\rho)=|j{\rangle}{\langle} j|\rho|j{\rangle}{\langle} j|$. The actually performed map
depends on the state preparation $\rho_{\sf P}$ that Alice choses for
${\sf P}$, and any (pure or mixed) state will be equivalent to a set of
probabilities $p_j^{(b)}={\langle}j|\rho_{\sf P}^{(b)}|j{\rangle}$ for the secret
parameter $j$ as follows 
\begin{equation}
\operatorname{Tr}_{{\sf P}}[\mathbb{M}^{(b)}(\, |\varphi{\rangle}{\langle}\varphi|\otimes\rho_{\sf P}^{(b)})]=
\sum_j{\mathrm{M}}_j^{(b)}(|\varphi{\rangle}{\langle}\varphi|)
{\langle}j|\rho_{\sf P}^{(b)}|j{\rangle}.
\end{equation}

\subsection{Reduction to trace-preserving maps} 
The maps ${\mathrm{M}}_j^{(b)}$ are generally trace-decreasing, i. e.  they
may be achieved with nonunit probability. In terms of the Kraus
decomposition for any input state $\rho$
\begin{equation}
{\mathrm{M}}_j^{(b)}(\rho)=\sum_i
E_{ji}^{(b)}\rho E_{ji}^{(b)}{}^\dag,\label{Kraus}
\end{equation}
this means that generally
\begin{equation}
\sum_i E^{(b)}_{ji}{}^\dag E^{(b)}_{ji}\le I.\label{decreas}
\end{equation}
Strictly trace-decreasing maps correspond to {\em aborting protocols},
namely when Alice doesn't succeed in achieving the map the protocol is
aborted. By completing the sum in Eq. (\ref{decreas}) with
additional terms in order to get the identity, we see that a trace
decreasing map is equivalent to a trace preserving one, with
additional outcomes $i$ corresponding to the protocol aborted.

\subsection{Reduction to unitary} 
Alice has unlimited technology, 
whence she can always achieve $E_{ji}^{(b)}$ {\em knowingly}, i. e.  she
has the option of achieving each trace-preserving map
${\mathrm{M}}_j^{(b)}$ as a {\em perfect pure measurement}. This can be
done as follows (in the following we will temporarily drop the indices
$b$ and $j$). The trace-preserving QO can be written in the form 
\begin{equation}
\begin{split}
{\mathrm{M}}(\rho)&=\operatorname{Tr}_{\sf F} [E\rho E^\dag],\\
E&=\sum_i E_i\otimes |i{\rangle}_{\sf F}\in{\sf B}({\sf H},{\sf K}\otimes{\sf F})
\mbox{ isometry},
\end{split}
\end{equation}
for a suitable ancillary space ${\sf F}$ (notice the tensor notation
$E\otimes|\psi{\rangle}_{\sf F}$ that for $E$ operator in ${\sf B}({\sf H},{\sf K})$ 
represents an ``extension'' operator from ${\sf H}$ to 
${\sf K}\otimes{\sf F}$.  By unitary embedding ${\sf H}$
into ${\sf K}\otimes{\sf F}\simeq{\sf H}\otimes{\sf A}$ for another suitable ancillary
space ${\sf A}$ as $E=U (I_{\sf H}\otimes |\omega{\rangle}_{\sf A})$, with $U$ unitary on
${\sf H}\otimes{\sf A}\simeq{\sf K}\otimes{\sf F}$, we have
\begin{equation}
{\mathrm{M}}(\rho)=\operatorname{Tr}_{\sf F} [U (\rho\otimes |\omega{\rangle}{\langle}\omega|_{\sf A}) U^\dag],
\end{equation}
namely Alice prepares the ancilla (and decoy systems) in the state
$|\omega{\rangle}$, and then performs a complete von Neumann measurement on  
${\sf F}$, with outcome $i$, which she keeps secret [the possibility of
using a more general type of measurement is already considered in an
extended space ${\sf F}$]. The strictly trace-decreasing case would
correspond to write 
\begin{equation}
{\mathrm{M}}(\rho)= 
\operatorname{Tr}_{\sf F} [(I_{\sf K}\otimes\Sigma_{\sf F}) U (\rho\otimes |\omega{\rangle}{\langle}\omega|_{\sf A})
U^\dag],
\end{equation}
with $\Sigma_{\sf F}$  orthogonal projector on a subspace
of ${\sf F}$. In conclusion, Alice can achieve the QO 
${\mathrm{M}}(\rho)=\sum_i E_i\rho E_i^\dag$ {\em knowingly} by: 
(1) preparing an ancilla/decoy state $|\omega{\rangle}_{\sf A}\in{\sf A}$; 
(2) performing a unitary transformation $U$ on ${\sf H}\otimes{\sf A}$; 
(3) performing a complete von Neumann measurement on ${\sf F}$, with
${\sf K}\otimes{\sf F}\simeq{\sf H}\otimes{\sf A}$ and outcome $i$; (4) sending ${\sf K}$ to Bob. 
Notice that we can have both extending and restricting protocols,
depending on the choice of ${\sf A}$ and ${\sf F}$. At this point, the
encoding maps are given by  
\begin{equation}
\begin{split}
&{\mathrm{M}}^{(b)}(|\varphi{\rangle}{\langle}\varphi|)=\\&
\!\!\!\sum_j p_j^{(b)}
\operatorname{Tr}_{{\sf F}}[(I_{\sf K}\otimes\Sigma_{j{\sf F}}^{(b)}) 
U_j^{(b)}(|\varphi{\rangle}{\langle}\varphi|\otimes|\omega{\rangle}{\langle}\omega|_{\sf A})U_j^{(b)}{}^\dag],\label{mmap0}
\end{split}
\end{equation}
with $|\omega{\rangle}$ independent on $j$ and $b$, since any dependence can
be included in $U_j^{(b)}$. Notice that if all orthogonal projectors
$\Sigma_j^{(b)}$ on subspaces of ${\sf F}$ have the same rank, their
dependence on $j$ and $b$ can also be included in $U_j^{(b)}$, but
generally $\operatorname{rank}(\Sigma_j^{(b)})$ depends on both $j$ 
and $b$. For the moment, we will focus attention on 
the case in which $\operatorname{rank}(\Sigma_j^{(b)})$ is independent on $j$. 
Now, by considering the unitary operator
$U^{(b)}=\sum_j U_j^{(b)}\otimes |j{\rangle}{\langle} j|$ over
${\sf H}\otimes{\sf A}\otimes{\sf P}\simeq{\sf K}\otimes{\sf F}\otimes{\sf P}$, we see that 
Alice can achieve any possible commitment step as follows
\begin{equation}
\begin{split}
&{\mathrm{M}}^{(b)}(|\varphi{\rangle}{\langle}\varphi|)=\\
&\!\!\!
\operatorname{Tr}_{{\sf F}\otimes{\sf P}}[(I_{\sf K}\otimes\Sigma_{\sf F}^{(b)}) 
U^{(b)}(|\varphi{\rangle}{\langle}\varphi|\otimes|\omega{\rangle}{\langle}\omega|_{\sf A}\otimes\rho_{\sf P})
U^{(b)}{}^\dag],\label{mmap}
\end{split}
\end{equation}
where also $\rho_{\sf P}$ is independent on $b$, whence, since also 
the probabilities $p_j^{(b)}$ will be independent on $b$, we will
denote them simply as $p_j$.

\subsection{Opening step}
In a protocol which is completely and perfectly verifiable Alice 
tells $b$ along with the {\em secret parameter $j$} and the 
{\em secret outcome $i$} to Bob, who verifies the state 
$E_{ji}^{(b)}|\varphi{\rangle}$. (In a protocol that is not
perfectly verifiable, the disclosed state is generally mixed, e. g. 
Alice keeps the outcome $i$ secret, or join outcomes in 
composite events as in a degenerate Luders measurement).
However, we emphasize that, whatever is the rule for the opening,
Alice has always the option of achieving the encoding QO by 
performing a complete von Neuman measurement. 
Since the local QO's on ${\sf K}$ and ${\sf F}\otimes{\sf P}$ commute, 
Alice has the possibility of: (1) first sending ${\sf K}$ to Bob; (2) then
performing the measurement on ${\sf F}\otimes{\sf P}$ at the very last moment
of the opening. As we will see, this is the basis for Alice EPR
cheating attacks. Notice that strictly trace-decreasing QO's---i. e. 
aborting protocols---pose limitations to Alice's EPR cheating. In
fact, Alice cannot delay the abortion of the protocol at the opening,
and must declare it at the commitment. Since both secret parameters
$j$ and $i$ can be conveniently measured by Alice, they can be treated
on equal footings as a single parameter $J\equiv(j,i)$.  The two maps
are then 
\begin{equation}
\begin{split}
{\mathrm{M}}^{(b)}(|\varphi{\rangle}{\langle}\varphi|)=&
\sum_j p_j{\mathrm{M}}_j^{(b)}(|\varphi{\rangle}{\langle}\varphi|)\\=&
\sum_J E_J^{(b)} |\varphi{\rangle}{\langle}\varphi|E_J^{(b)}{}^\dag,
\end{split}
\end{equation}
with $E_J^{(b)}\doteq\sqrt{p_j}E_{ji}^{(b)}\in{\sf B}({\sf H},{\sf K})$.

\subsection{Reduction to a single commitment step}
A protocol with more than a commitment step generally consists of a
sequence of conditioned QO's, namely in which one party is requested
to make a different  QO, say $\{\mathrm{N}^{(x)}\}$, depending on the
outcome $x$ of a previous QO from the other party.  However, the same
result is achieved by automatizing the conditioned QO, and using
instead the unconditioned one $\mathbb{N}=\sum_x \mathrm{N}^{(x)}\otimes
\mathrm{P}_x$ on an extended Hilbert space ${\sf H}\otimes{\sf N}$, without even
knowing $x$. If the knowledge of $x$ is requested only at the opening--as
for nonaborting protocols---then the orthogonal measurement
$\mathrm{P}_x$ can be delayed up to the opening 
moment, since the {\em notepad} space ${\sf N}$ is kept by the party.
Then, analogously as for a single commitment step, each QO can be achieved knowingly, by means of a pure measurement,
with a suitable unitary embedding. Again, since the
measurement-ancillary space (${\sf F}$ in the above analyzed single
commitment step) is kept by the considered party, its measurement can
be delayed up to the opening moment (for strictly trace decreasing
maps the two parties can agree to declare the abortion at 
the end of the whole commitment). At this point, we have a sequence
of interlaced unitary operators, one from each party alternatively,
e. g.  for three steps ${U'}_A^{(b)}U_B U_A^{(b)}$, where clearly the
unitary transformation by Bob $U_B$ cannot depend on $b$. Now, for a 
numerable set of possible unitary transformations $U_B\in\{U_l\}$, Bob
can use instead the unitary $\mathbb{U}_B=\sum_l 
U_l\otimes|l{\rangle}{\langle} l|$ by preparing a state from an orthogonal set
$\{|l{\rangle}\}$ on an additional Hilbert space. Therefore, the choice of
the unitary is equivalent to the state preparation of another
anonymous-state Hilbert space. In conclusion, from the arguments above
we see that the whole multi-step (non aborting) protocol is equivalent to a
single-step one, with larger spaces ${\sf H}$, ${\sf K}$, ${\sf A}$, ${\sf F}$, and
${\sf P}$. We don't know what is the minimal Hilbert space for 
anonymous-state preparation of a generally  continuous set of 
unitary operators, for which one may need a non separable
space. Notice that with a teleportation protocol it is possible to
achieve any contraction on a space ${\sf H}$ by performing a state 
preparation on the space ${\sf H}\otimes{\sf H}$ of the entangled 
resource, however, only with probability equal to $\operatorname{dim}({\sf H})^{-2}$.

\subsection{Classical protocols} It is obvious that a classification of
quantum protocols must include also the classical ones as a particular
case. In fact, the classical protocols correspond to consider just
orthogonal states, and QO's on abelian operator algebras. Consider,
for example, a  one-way trapdoor function $f_A(j)$, where the integer
$j$ plays the role of the secret parameter. Let the value $b$ if the
committed bit be the parity of $j$. Then, Alice sends the state  
$|n{\rangle}\otimes|f_A{\rangle}$ to Bob, with $n=f_A(j)$. Bob can verify that $f_A$
is indeed one-way. However, since he cannot compute $j$ from the
knowledge of $n$, he can just guess whether $j$ is even or odd. At the
opening Alice tells $j$ to Bob, and Bob verifies that indeed
$n=f_A(j)$ and evaluates the parity of $j$.  

\section{Cheating}
For the moment we will focus attention on non-aborting protocols,
postponing the discussion of the aborting (strictly decreasing)
case. Alice can cheat at two different moments: 
before and after the commitment. We will name the two cases:  {\em
pre-cheating} and {\em post-cheating}, respectively. Bob, as we will
see, can perform a combined attack before and after the commitment.
The possibility also for Alice of performing a combined attack will be
also discussed.

\subsection{Alice pre-cheating attacks} These correspond to prepare the
ancillary spaces ${\sf A}\otimes{\sf P}$ in a state not of the prescribed form
$|\omega{\rangle}{\langle}\omega|_{\sf A}\otimes\rho_{\sf P}$. This will generally lead to QO's different
from the ones prescribed from the protocol. 
In the following, we will not further analyze pre-cheating, for the
two following reasons: (1) it seems that there is no practical use
for Alice to cheat before knowing if the committed bit needed to 
be changed;
(2) either there is a chance that the pre-cheating will be detected at
the opening, or it would lead to QO's indistinguishable from the
prescribed ones, in which case it will give the same result of a
post-cheating attack considered in the following. Finally, notice that
a cheating attack based on changing the prescribed unitaries $U^{(b)}$
belongs to the same class of {\em pre-cheating} attacks, and the same
considerations hold. 

\subsection{Alice post-cheating attacks} After the commitment and before the
opening Alice can try to cheat by performing a unitary transformation
$V$ on ${\sf F}\otimes{\sf P}$: this is the so-called EPR attack. The maneuver
will not change the QO's ${\mathrm{M}}^{(b)}$, however, it will change the
Kraus decompositions---which are relevant at the
opening---giving a new set of contractions
$\{E_J^{(b)}\}\rightarrow\{E_J^{(b)}(V)\}$ with the same cardinality, in
the following way  
\begin{equation}
E_J^{(b)}(V)=\sum_L E_L^{(b)} V_{JL},\qquad V_{JL}= {\langle}J |V|L{\rangle}.
\end{equation}
Another attack available to Alice is also that of declaring a $J$ 
different from the actual outcome: however, since Alice doesn't 
know the anonymous state $|\varphi{\rangle}$, she must adopt a fixed 
rule to scramble the $J$'s, and being just a permutation, this 
cheating is again equivalent to a unitary cheating transformation 
$V$.
\par The probability that Alice can cheat successfully in 
pretending having committed, say, $b=1$, whereas she 
committed $b=0$ instead, is given by
\begin{equation}
P_c^A(V,\varphi)=\sum_J 
\frac{|{\langle}\varphi|E_J^{(0)}{}^\dag(V)
E_J^{(1)}|\varphi{\rangle}|^2}{|\!|E_J^{(1)}\varphi|\!|^2},\label{PcAV}
\end{equation}
and it clearly depends on the anonymous state $|\varphi{\rangle}$ and on
the cheating transformation $V$. However, Alice doesn't know 
$|\varphi{\rangle}$, and the optimal choice of $V$ obviously depends on 
$|\varphi{\rangle}$. So, which is the transformation $V$ to be used? 
Without any knowledge of $|\varphi{\rangle}$, the best Alice can do is to 
adopt a conservative strategy, by choosing the $V$ such that the
minimum $P(V,\varphi)$ for $|\varphi{\rangle}$ chosen by Bob is 
maximum, namely she maximizes her probability of cheating in 
the worst case, corresponding to the {\em minimax} choice of $V$
\begin{equation}
(P_c^A)_\mu\doteq\max_V\min_\varphi 
P_c^A(V,\varphi).\label{minimaxP}
\end{equation}
It is evident that in this way a game theoretical situation arises, in
which Bob choses $|\varphi{\rangle}$ and Alice choses $V$, with the 
probability $P(V,\varphi)$ playing the role of a {\em payoff 
matrix}. Obviously Alice and Bob can generally adopt randomized 
strategies, which can then be purified via entanglement with an 
ancillary system. However, in the general situation the game is 
further complicated by the fact that Bob's choice for $|\varphi{\rangle}$ is 
also dictated by maximization of his own probability of 
cheating (see later), and all other parameters---such as Alice 
secret parameter $j$---must enter the game. Since we 
are only interested to set the debate on the impossibility 
proof\cite{MayersQBC97,LoChauQBC97,LoQBC97} via a 
complete classification of all protocols and cheating attacks,
this game situation, which arises as a consequence of using 
anonymous states, will be analyzed elsewhere.
\par Another possibility for Alice's choice of $V$ would be that of 
maximizing the probability $P(V,\varphi)$ averaged over all 
anonymous states, with the unitarily invariant probability measure
$\operatorname{d}\mu(\varphi)$ on the (compact) manifold of pure states,
namely
\begin{equation}
(P_c^A)_{av}=\max_V \int\operatorname{d}\mu(\varphi)P_c^A(V,\varphi).
\label{PcA}
\end{equation}
However, such a (pure) strategy will not be optimal if Bob chooses 
a non uniform probability distribution, e. g.  a delta-function, and 
the actual probability of cheating could be much lower than the
one in Eq. (\ref{PcA}). It is obvious that for compact 
manifold of states---i. e.  for finite dimensions---then the two
probabilities in Eqs. (\ref{minimaxP}) and (\ref{PcA}) will be 
related by a constant depending on the dimension of ${\sf H}$. \par 
The evaluation of the average in Eq. (\ref{PcA}) is made difficult 
by the presence 
of the norm in the denominator. When the encoding for $b=1$
is random unitary, i. e.  $E_J^{(1)}=\sqrt{p_J}U_J^{(1)}$ with
unitary $U_J^{(1)}$, the evaluation of the average in Eq. 
(\ref{PcA}) is simplified by the following identity which holds for 
any couple of operators $A,B\in{\sf B}(H)$ for 
$d\doteq\operatorname{dim}({\sf H})<\infty$ 
\begin{equation}
\begin{split}
&\int\operatorname{d}\mu(\varphi){\langle}\varphi|A|\varphi{\rangle}{\langle}\varphi|B|\varphi{\rangle}\\&=
\frac{1}{d(d+1)}[\operatorname{Tr}(A)\operatorname{Tr}(B)+\operatorname{Tr}(AB)].\label{idAV}
\end{split}
\end{equation}
Using Eq. (\ref{idAV}) the averaged probability in Eq. (\ref{PcA})
rewrites
\begin{equation}
\begin{split}
&(P_c^A)_{av}=\frac{1}{d+1}\\&+\frac{1}{d(d+1)}
\max_V \sum_J\left|\sum_L\operatorname{Tr}\left(U_J^{(1)}{}^\dag 
E_L^{(0)}\right)V_{LJ}\right|^2,
\end{split}
\end{equation}
which can be bounded as follows
\begin{equation}
\begin{split}
\frac{1}{d+1}\le
(P_c^A)_{av}&\le\frac{1}{d+1}+\frac{1}{d(d+1)}
\left|\!\left|{\mathbb Z}\right|\!\right|_1,\\ {\mathbb Z}_{(JL)K}&=
\operatorname{Tr}[U_K^{(1)}{}^\dag E_J^{(0)}]\operatorname{Tr}[U_K^{(1)} E_L^{(0)}{}^\dag],\label{boundsA}
\end{split}
\end{equation}
where $\left|\!\left|\cdot\right|\!\right|_1$ denotes the trace-norm, and the
matrix ${\mathbb Z}_{(JL)K}$ has to be considered as rectangular, with
$(JL)$ as a single index. From Eq. (\ref{boundsA}) we see that in
order to reduce Alice's cheating probability we better increase the
dimension of the anonymous-state Hilbert space. The upper 
bound in Eq. (\ref{boundsA}) could be useful for proving 
unconditional security of the protocol: however, we don't know if 
the trace norm in Eq. (\ref{boundsA}) is bounded as 
$\left|\!\left|{\mathbb Z}\right|\!\right|_1\le d^2$, otherwise the bound 
(\ref{boundsA}) would be useless (one can check that 
$\left|\!\left|{\mathbb Z}\right|\!\right|_1=d^2$ in the perfectly concealing case).

\subsection{Bob cheating} Bob can try to cheat by making the {\em best
discrimination} between the two maps ${\mathrm{M}}^{(b)}=\sum_j
p_j{\mathrm{M}}_j^{(b)}$. However, since he doesn't know the probabilities 
$p_j$ actually used by Alice, his strategy will be suboptimal,
and his actual cheating probability $P_c^B$ will be lower than 
the probability $(P_c^B)_{opt}$ corresponding to the optimal 
strategy with the right probabilities 
$p_j$. Since map-discrimination is generally more reliable with
the map acting locally on an entangled 
state\cite{entang_meas}, instead of preparing $|\varphi{\rangle}\in{\sf H}$ Bob  
prepares an entangled state on ${\sf H}\otimes{\sf R}$ and sends only ${\sf H}$ to
Alice. (Here, we can see clearly that the use of anonymous states in the
protocol, while limiting Alice EPR cheating attacks, at the same time
allows Bob to perform EPR attacks himself). Therefore, Bob's optimal
probability of cheating is bounded as follows (for equally probable
bit values $b=0,1$)  
\begin{equation}
\begin{split}
&P_c^{B}\le(P_c^{B})_{opt}=\frac{1}{2}+
\\&\!\!\!
\max_{|\varphi{\rangle}\in{\sf H}\otimes{\sf R}}\frac{1}{4}
\left|\!\left|{\mathrm{M}}^{(1)}\otimes\mathrm{I}_{\sf R} (|\varphi{\rangle}{\langle}\varphi|)-
{\mathrm{M}}^{(0)}\otimes\mathrm{I}_{\sf R}
(|\varphi{\rangle}{\langle}\varphi|)\right|\!\right|_1\\
&=\frac{1}{2}+\frac{1}{4}\left|\!\left|{\mathrm{M}}^{(1)}-{\mathrm{M}}^{(0)}
\right|\!\right|_{cb},\label{cbcheat}
\end{split}
\end{equation}
where $\left|\!\left|\cdot\right|\!\right|_{cb}$ denotes the completely bounded (CB) 
norm\cite{Paulsen}, and we used the fact that the difference of two CP
maps is Hermitian-preserving, whence its CB-norm is achieved on a 
normalized vector in ${\sf H}\otimes{\sf R}$.
Notice that for trace-preserving QO's the difference
${\mathrm{M}}^{(1)}-{\mathrm{M}}^{(0)}$ is 
never completely positive, and generally an entangled anonymous state
improves the discrimination, whereas for aborting protocols the QO's
are strictly trace-decreasing, and the difference map can be completely
positive itself, in which case the EPR attack is of no use (for such
analysis on discriminations between QO's, see Ref. \cite{entang_meas}).
\begin{table}[hbt]\begin{center}\begin{tabular}{|l|l|}
\hline\hline
Symbol&Hilbert space \\ \hline\hline
${\sf H}$ & Anonymous state \\ \hline
${\sf K}$ & Output \\ \hline
${\sf A}$ & Preparation ancilla/decoy \\ \hline
${\sf P}$ & Secret parameter \\ \hline
${\sf F}$ & Measurement ancilla \\ \hline
${\sf R}$ & Bob cheating space \\ \hline
${\sf Rng}(\Sigma)$ & Range of $\Sigma$ (abortion)
\\ \hline \hline
\end{tabular}\end{center}
\caption{List of Hilbert spaces needed for protocol and cheating
attacks classification.}
\label{t:spacesa}
\end{table}
\begin{table}[hbt]\begin{center}\begin{tabular}{|l|l|l|l|}
\hline\hline
&start & commitment & after commitment\\
\hline\hline
Alice & ${\sf A},{\sf P}$ & ${\sf H},{\sf A},{\sf P}$ & ${\sf F},{\sf P}$\\ \hline
Bob & ${\sf H},{\sf R}$ & ${\sf R}$ & ${\sf K},{\sf R}$\\ \hline \hline
\end{tabular}\end{center}
\caption{Who owns which space and when.}\label{t:spacesb}
\end{table}
\subsection{Discussion on the aborting (strictly trace-decreasing) protocols}
In the simplest case in which the projector $\Sigma$ is independent on
both $b$ and $j$, Alice can launch an EPR attack easily, performing it on
the range space of $\Sigma$. However, when the rank of $\Sigma$
depends on $b$, an EPR attack has a probability of being detected
by Bob at the opening when the attack leads to a larger Kraus
cardinality than the prescribed one. Notice also that, in general, a
dependence of $\operatorname{rank}(\Sigma)$ on $b$ and/or $j$ will enhance Bob's
probability of cheating.  

Up to now we have seen that in order to classify all possible
protocols and cheating attacks we need to consider seven Hilbert
spaces with different physical meanings: these are summarized in Tables
\ref{t:spacesa} and \ref{t:spacesb}.

\subsection{Perfectly concealing protocols}
A perfectly concealing protocol
means that the CB-norm in Eq. (\ref{cbcheat}) is 
zero, namely the two maps are the same.
Therefore, the two Kraus are connected via a unitary transformation
$V$ on ${\sf F}\otimes{\sf P}$, and Alice can cheat with probability one,
namely the protocol {\em is not binding}. 

\subsection{Approximate concealing protocols} We consider now the case in which 
Bob's probability of cheating for the optimal strategy is
infinitesimally close to the pure guessing probability $\frac{1}{2}$,
which means that also the CB-norm distance between the maps is
infinitesimal, i. e.  
$|\!|{\mathrm{M}}^{(1)}-{\mathrm{M}}^{(0)}|\!|_{cb}=\varepsilon$. We 
emphasize that generally $\varepsilon$ is vanishing for increasing
dimension of ${\sf K}$ (see, for example, some protocols given by 
Yuen\cite{yuenqbc4}, where the approximately concealing condition is achieved for increasingly large number of decoy systems), 
and no obvious continuity argument can be invoked to assert that Alice
cheating probability will approach unit for vanishing
$\varepsilon$. More precisely, in the present context based on
anonymous states, such an argument (which is at the basis of 
the impossibility proof of
Refs.\cite{MayersQBC97,LoChauQBC97,LoQBC97})
would imply that for both the minimax and the averaged
strategies in Eqs. (\ref{minimaxP}) and (\ref{PcA}) Alice 
probability of cheating would be infinitesimally close to unit for 
$\varepsilon\to0$, namely 
\begin{equation}
1-(P_c^A)_{\mu,av}=\omega
\left(\left|\!\left|{\mathrm{M}}^{(1)}-{\mathrm{M}}^{(0)}\right|\!\right|_{cb}\right),
\label{impossibility}
\end{equation}
for some function $\omega(\varepsilon)$ independent on the 
dimension of ${\sf K}$ and vanishing with $\varepsilon$.
However, using anonymous states such assertion may 
turn out to be false. In fact, it is obvious that if there is an 
alternate Kraus 
decomposition $\{E_J^{(0)}(V)\}$ for the map ${\mathrm{M}}^{(0)}$
such that the two Kraus $\{E_J^{(0)}(V)\}$ and $\{E_J^{(1)}\}$ are
close, then the protocol is approximately concealing and not 
binding, since (see Appendix)
\begin{eqnarray}
&&(P_c^{B})_{opt}-\frac{1}{2}= 
\frac{1}{4}\left|\!\left|{\mathrm{M}}^{(1)}-{\mathrm{M}}^{(0)}\right|\!\right|_{cb}
\nonumber\\ &&\le
\frac{1}{2}\sqrt{\left|\!\left|\sum_J\left|E_J^{(0)}(V)-E_J^{(1)}\right|^2
\right|\!\right|},
\label{boundbob}\\
&&\!\!\!P_c^A(V,\varphi)\ge\left[1-\frac{1}{2} 
\left|\!\left|\sum_J\left|E_J^{(0)}(V)-E_J^{(1)}\right|^2\right|\!\right|\right]^2,
\label{boundalic}
\end{eqnarray}
where $\left|\!\left|\cdot\right|\!\right|$ denotes the usual spectral norm, 
and for any operator $A$ we use the customary abbreviation 
$|A|^2\doteq A^\dag A$.
However, the impossibility proof would be true if a bound of the 
form (\ref{boundbob}) would be satisfied in the reverse direction,
in which case one would have
\begin{equation}
\begin{split}
1-(P_c^A)_{\mu,av}&\le\min_V
\left|\!\left|\sum_J\left|E_J^{(0)}(V)-E_J^{(1)}\right|^2\right|\!\right|\\&\le
\omega\left(\left|\!\left|{\mathrm{M}}^{(1)}-{\mathrm{M}}^{(0)}
\right|\!\right|_{cb}\right),
\label{endofstory}
\end{split}
\end{equation}
which would correspond to the following {\em continuity 
argument}: if two CP maps are close in CB-norm, then
for a given fixed Kraus decomposition for one of the two maps, 
there is always an alternate Kraus decomposition for the other 
map such that the two are close. Since one also has that
$|\!|A|\!|\le|\!|A|\!|_2$,where $\left|\!\left|\cdot\right|\!\right|_2$ now denotes the 
Frobenius (Hilbert-Schmidt) norm, the bounding (\ref{endofstory})
could also be written with the Frobenius norm in 
the middle term (see Eqs. (\ref{boundbob}) and 
(\ref{boundalic})), in 
which case the minimum over $V$ would be in the form of 
a Procrustes problem \cite{Procrustes}. \par 
Since as regards the cheating probabilities we have considered 
only the case of non-aborting protocols with perfect-verification, 
proving the continuity argument (\ref{endofstory}) or directly the
bound (\ref{impossibility}) would means that a secure protocol 
can still be searched outside such class of protocols. On the other 
hand, finding a counterexample to Eq. (\ref{impossibility})
would provide a perfectly verifiable and unconditionally secure 
protocol. 
\par Finally, a few words on the possibility of a combined
pre/post-cheating Alice attack. It is clear that if it leads to a set
of QO's different from the prescribed ones, then it can be detected by
Bob at the opening (if it gives the prescribed set of maps,
then the same effect can be better achieved by post-cheating). 
However, in principle, it may help in increasing the overall Alice's
cheating probability, particularly when there is possibility of
abortion,  i. e.  for strictly trace-decreasing protocols.  

\section{Discussion}
The discrepancy between the previous analysis and the analysis
beneath the impossibility proof 
\cite{MayersQBC97,LoChauQBC97,LoQBC97}
is essentially due to the fact that the latter is based on the
assumption that the starting state of the commitment protocol
is openly known, in the sense that the probability distribution of
the state is given, and then the corresponding mixed state can be 
purified. The general underlying idea is that the protocol should
be processed by {\em machines}, and therefore all probability 
distributions are defined, and purified inside the machines. 
However, such an assumption is certainly not realistic for a 
cryptographic protocol, where each party has actually the 
freedom of changing or tuning the machine, namely chosing 
any desired probability distribution. Or else, one would need to 
purify the human being himself. Now, if a parameter is secret for 
a party, nobody forbids him/her to believe that the other party is 
still using the same prepared machine, and accordingly to adopt
a Bayesian approach with the known a priori probability 
distribution. However, in practice, the other party could have
used another machine and/or with a different preparation. 
One can continue to argue on this line, asserting that 
changing the machine is equivalent to use a larger machine, or, 
in other words, that an unknown probability distribution can be 
regarded as an a priori uniform distribution on the space of 
probability distributions. This line of reasoning, however,
constitutes a very dangerous argument in a proof, since it is 
equivalent to consider {\em infinite machines} or, 
equivalently, uniform probabilities on infinite sets, which then 
must vanish everywhere. In addition, for infinite probability spaces
one needs infinite dimensional purifications, (even worst, for
continuous spaces one may need non separable Hilbert spaces). This
would invalidate an impossibility proof based on a non proved
continuity argument which {\em a fortiori} must apply to infinite  
dimensions. Finally, one could now argue that in the real world the
machines must be bounded: however, this assertion would 
contradict not only the previous assumption of uniform 
probabilities (otherwise, which non uniform probabilities are to be
adopted?), but also the fact that the proof is purported for 
{\em unconditionally security}, with both Alice and Bob supposed 
to possess unlimited technology.
\par The above hill-posed mathematical framework arises from the
Bayesian approach to secret parameters, dictated from the {\em closed
system} modeling with fixed machines and purification of
probabilities. This model, along with the assumed unbounded technology
for both parties, necessarily lead to infinities which don't allow
unproven continuity arguments, thus  falsifying the proof. 
Alternative to the previous approach, we have the realistic
{\em finite open system} approach, in which unknown parameters are
treated as such, without the need of any a priori probability
distribution, in which we can address the problem for finite dimension
with the parameter $\varepsilon$ depending on it. As well known, 
the need of treating unknown parameters without a priori probability
distribution is the reason why in detection and estimation 
theory\cite{Holevo,Helstrom} we have both the minimax and the Bayesian
approaches. Then, if one  
proceeds by treating unknown parameters as such, no openly 
known state can be assumed, and the anonymous state encoding 
of Yuen\cite{yuenqbc3,yuenqbc4} leads to the present 
classification of protocols. Notice that if the initial state 
$|\varphi{\rangle}$ is openly known, then for that given fixed states
all QO's can be regarded as random unitary 
transformations (since all states are connected by unitary
transformations), and this lead to the simple form of Alice 
cheating probability in terms of 
fidelities\cite{MayersQBC97,LoChauQBC97,LoQBC97},
whereas in the present context the probability of cheating has the
more involved form (\ref{PcAV}), due to the fact that the state 
$|\varphi{\rangle}$ is unknown, and that there are QO's that don't admit 
random unitary Kraus decompositions.
\par Finally, a few words on the possibility of aborting protocols.
This possibility was not considered in Refs. 
\cite{MayersQBC97,LoChauQBC97,LoQBC97}, since also this
in practice arises as a consequence of not assuming
openly known probabilities. In fact, in a closed model of 
interacting machines with purified parameters, every 
transformation would be unitary. However, one could reasonably
argue that if the protocol aborts, then another protocol must be
started, and the procedure will be repeated as long as the bit 
is not successfully committed, and that such a succession of 
protocols is itself a non aborting protocol. Such kind of protocols
that could be chained {\em ad infinitum} can be regarded as infinite 
convex combinations of protocols on infinite dimensional 
anonymous spaces ${\sf H}$, (the QO will be trace-preserving only
for infinite dimensions). Again one can see that a {\em closed system}
approach necessarily lead to infinite dimensions.
\appendix
\section{Derivation of the bounds (\ref{boundbob}) and (\ref{boundalic}) for cheating probabilities}
Using Jensen inequality as suggested in Ref. \cite{yuenqbc4},
the Alice's cheating probability can be bounded from below  as follows
[here we use the abbreviate notation $F_J^{(0)}\doteq
E_J^{(0)}(V)$]
\begin{equation}
\begin{split}
&P_c^A(V,\varphi)=\sum_J |\!|E_J^{(1)}\varphi|\!|^2
\left(\frac{|{\langle}\varphi|F_J^{(0)}{}^\dag 
E_J^{(1)}|\varphi{\rangle}|}{|\!|E_J^{(1)}\varphi|\!|^2}\right)^2\\&\ge \left|\sum_J
{\langle}\varphi|F_J^{(0)}{}^\dag E_J^{(1)}|\varphi{\rangle}\right|^2
\ge \left(\Re\sum_J {\langle}\varphi|F_J^{(0)}{}^\dag E_J^{(1)}
|\varphi{\rangle}\right)^2\\ &=
\left[1-\frac{1}{2}{\langle}\varphi|\sum_J\left|F_J^{(0)}-E_J^{(1)}\right|^2
|\varphi{\rangle}\right]^2\\ &\ge
\left[1-\frac{1}{2}\left|\!\left|\sum_J\left|F_J^{(0)}-E_J^{(1)}\right|^2
\right|\!\right|\right]^2,
\end{split}
\end{equation}
where we used $|A|^2\doteq A^\dag A$, and the fact that for 
$P\ge 0$ one has $|\!|P|\!|=\sup_{|\!|\varphi|\!|=1}
{\langle}\varphi|P|\varphi{\rangle}$.
On the other hand, we have that 
\begin{equation}
\begin{split}
&\left|\!\left|{\mathrm{M}}^{(0)}-{\mathrm{M}}^{(1)}\right|\!\right|_{cb}^2=
\left|\!\left|\operatorname{Tr}_{\sf F} [F^{(0)}\cdot F^{(0)}{}^\dag-E^{(1)}\cdot
E^{(1)}{}^\dag]\right|\!\right|_{cb}^2\\&\le\left|\!\left|
\operatorname{Tr}_{\sf F}\right|\!\right|_{cb}
\left|\!\left|F^{(0)}\cdot F^{(0)}{}^\dag-E^{(1)}\cdot
E^{(1)}{}^\dag\right|\!\right|_{cb}^2\\&\le
\left|\!\left|F^{(0)}\cdot F^{(0)}{}^\dag-E^{(1)}\cdot
E^{(1)}{}^\dag\right|\!\right|_{cb}^2\\
&\le \left[\left|\!\left|
I\cdot F^{(0)}{}^\dag\right|\!\right|_{cb}
\left|\!\left|F^{(0)}\cdot I-E^{(1)}\cdot I\right|\!\right|_{cb}
\right.\\&+\left.
\left|\!\left| F^{(0)}\cdot I\right|\!\right|_{cb}
\left|\!\left|I\cdot F^{(0)}{}^\dag-I\cdot
E^{(1)}{}^\dag\right|\!\right|_{cb}\right]^2\\ &
\le 4\left|\!\left|F^{(0)}-E^{(1)}\right|\!\right|^2\le4
\sup_{|\!|\varphi|\!|=1}{\langle}\varphi|\sum_J\left|F_J^{(0)}-E_J^{(1)}\right|^2
|\varphi{\rangle}\\&=4
\left|\!\left|\sum_J\left|F_J^{(0)}-E_J^{(1)}\right|^2\right|\!\right|.
\end{split}
\end{equation}
\section*{Acknowledgments}
This work has been jointly founded by the EC under the program 
ATESIT (Contract No. IST-2000-29681) and by the Department 
of Defense Multidisciplinary University Research Initiative (MURI)
program administered by the Army Research Office under Grant 
No. DAAD19-00-1-0177.
The author acknowledges a private communication by H. K. Lo 
stimulating this work, and extensive and accurate analysis and 
criticisms by H. P. Yuen,  who motivated this work. The author 
also would like to thank R. Werner for suggestions for further 
improving the bounds (\ref{boundbob}) and (\ref{boundalic}),
and  C. Bennet for clarifying the general philosophy and attitude 
which is at the basis of the impossibility proof.
 
\end{document}